\documentclass[10pt, a4paper]{article}
\usepackage[final]{lrec2026} 

\usepackage{enumitem}
\usepackage{amsmath}

\title{Personality Anchoring for Social Simulation: Linking Personality, Social Behavior, and Interaction Success with LLM Agents}

\name{Vahid Sadiri Javadi \textsuperscript{$\spadesuit$}, Aksa Aksa \textsuperscript{$\spadesuit$}, Fryderyk Róg \textsuperscript{$\spadesuit$} \\ 
\textbf{\large Lucie Flek \textsuperscript{$\spadesuit$}, Johanne R. Trippas \textsuperscript{$\clubsuit$}}}

\address{University of Bonn, Conversational AI and Social Analytics (CAISA) Lab, Bonn, Germany \textsuperscript{$\spadesuit$}\\RMIT University, School of Computing Technologies, Melbourne, Australia  \textsuperscript{$\clubsuit$}\\}

\abstract{Social interactions are shaped by the interplay of dispositional traits and situational context, yet systematically investigating how personality configurations between individuals jointly influence social behavior across diverse social contexts remains methodologically challenging. We address this gap by introducing a simulation pipeline adapted from the CHARISMA framework, which employs well-known movie characters and public figures as psychologically grounded agents for multi-LLM social simulation using a method we term \textit{personality anchoring}. We present a large-scale empirical study examining how dyadic Agreeableness composition influences social interaction outcomes across 1,010 simulated conversations. Our results reveal a monotonic relationship between dyadic Agreeableness composition and shared goal achievement, with Homogeneous-Agreeable pairs achieving success 10 times the rate of Homogeneous-Disagreeable pairs (62\% vs. 6\%). Behavioral mediation analysis reveals that Agreeableness shapes goal achievement partially through cooperative strategy selection, though it continues to predict outcomes within the same dominant strategy, indicating pathways beyond observable conversational behavior. Robustness analyses confirm high consistency of results across repeated simulations (ICC = 0.89) and stable personality expression across diverse scenarios, validating personality anchoring as a viable operationalization strategy.
 \\ \newline  }

\begin{document}

\maketitleabstract

\section{Introduction}
\label{sec:intro}

Understanding how dispositional traits and situational context jointly shape social interaction outcomes is central to social psychology. Attribution theory provides a foundational framework for this inquiry, explaining how individuals infer the causes of behavior by distinguishing between dispositional and situational factors \cite{heider1958psychology, kelley1967attribution, weiner1986attributional}. Among dispositional factors, individual differences are commonly operationalized through the Big Five personality framework, within which Agreeableness has been identified as the dimension most closely tied to interpersonal conflict processes and outcomes \cite{jensen2001agreeableness}, with highly agreeable individuals preferring negotiation and compromise, while those low in Agreeableness tend toward competitive or coercive strategies \cite{graziano1996perceiving, wilmot2022agreeableness}. Yet investigating how different personality configurations between interacting individuals jointly shape social outcomes across diverse contexts remains methodologically challenging. Traditional experimental methods, while yielding important insights \cite{aronson1990methods}, face limitations in scalability, reproducibility, and the systematic manipulation of complex social variables \cite{nosek2022replicability, open2015estimating, wicherts2016degrees}.

Recent advances in large language models (LLMs) have created new opportunities for computational social psychology by enabling the simulation of open-ended social interactions at unprecedented scale \cite{park2023generative, zhou2024sotopia}. Prior work has demonstrated that LLMs can effectively simulate Big Five personality traits with behaviors that human observers rate as believable \cite{jiang2024personallm, serapio2025psychometric, javadi2025cinemetric}, and frameworks such as SOTOPIA \cite{zhou2024sotopia} have enabled systematic evaluation of social intelligence in LLM-based agents. Studies on personality-conditioned agents have further explored how traits influence cooperation in games \cite{qiu2025networkgames} and social media behavior in large-scale simulations \cite{yang2024oasis}.

However, several important gaps remain. First, most existing simulations examine personality within relatively narrow settings, such as bargaining tasks or cooperative games, without grounding scenarios in a validated taxonomy of everyday human goals. Second, personality is typically introduced through explicit trait prompting (e.g., \textit{``you are highly agreeable''}), which frames personality as an instruction rather than a naturally occurring behavioral tendency. Third, there is limited focus on \textit{dyadic personality composition}, i.e., how different configurations of personality traits between two interacting individuals shape joint outcomes across varied social contexts. Fourth, existing evaluations tend to emphasize overall task outcomes or persona fidelity, often overlooking the \textit{behavior strategies} through which personality influences social outcomes, i.e., the conversational mechanisms that mediate the personality--outcome relationship.

In this paper, we address these gaps by adapting the CHARISMA framework \cite{sadirijavadi2026charisma}, which employs well-known movie characters and public figures as psychologically grounded agents for multi-LLM social simulation. Rather than assigning abstract trait scores, CHARISMA leverages LLMs' embedded knowledge of characters' backstories and behavioral tendencies to simulate personality-consistent behavior. We term this approach \textit{personality anchoring}. We present a comprehensive empirical study examining how the dispositional trait Agreeableness influences social interaction outcomes. Agreeableness is operationalized through systematic dyadic pairings of characters with crowd-sourced Big Five profiles from the Personality Database (PDB)\footnote{\url{https://www.personality-database.com/}}
 across large-scale simulated conversations spanning seven social goal categories derived from a taxonomy of 135 human goals \cite{chulef2001hierarchical}. Our paper makes the following contributions:

\begin{enumerate}[leftmargin=*]
    \item We introduce a personality-driven simulation methodology that integrates personality through character-based anchoring, a structured taxonomy of human goals, and behavior strategies in conversational interaction, enabling systematic analysis of how dispositional traits and situational factors jointly shape social behavior in simulated interactions.
    
    \item We conduct a large-scale empirical analysis of how dyadic Agreeableness composition shapes social behavior across seven social goal categories, two difficulty levels, and multiple interaction models.
    
    \item We provide a behavioral mediation analysis examining whether and how conversational strategies mediate the relationship between personality composition and interaction outcomes, distinguishing between direct and indirect pathways of personality influence.
    
    \item We evaluate robustness along two dimensions: \textit{(i)} result consistency across repeated simulations and \textit{(ii)} personality expression stability across diverse scenarios, assessing the reliability of character-based anchoring as a personality operationalization strategy.
\end{enumerate}

The code, dataset, full list of behavior strategies and characters, and behavioral analysis scripts are publicly available.\footnote{\url{https://github.com/vahidsj/PersonalityAnchoring}}

\section{Related Work}
\label{sec:related-work}

\subsection{LLM-Based Social Simulation}
\label{sec:rw-simulation}

LLM-powered social simulation has scaled rapidly since the introduction of Generative Agents \cite{park2023generative}, which demonstrated that 25 LLM agents could sustain coherent social behavior, including relationship formation and activity coordination, over multiple simulated days using memory, reflection, and planning components. SOTOPIA \cite{zhou2024sotopia} shifted focus to systematic evaluation, introducing 90 social scenarios and a 7-dimensional evaluation framework assessing goal completion, relationship maintenance, and social norm adherence, with GPT-4 as an LLM-based evaluator. Follow-up work has extended this ecosystem: SOTOPIA-$\Omega$ \cite{zhang2025sotopia_omega} injects negotiation strategies enabling 7B models to surpass GPT-4 on social goals, while Sotopia-RL \cite{yu2025sotopia_rl} introduces utterance-level multi-dimensional rewards for training socially intelligent agents. At larger scales, OASIS \cite{yang2024oasis} simulates up to one million agents on social media platforms, replicating information spreading and group polarization dynamics. AgentSociety \cite{piao2025agentsociety} integrates Maslow's hierarchy of needs and the Theory of Planned Behavior into 10,000+ agents, successfully reproducing real-world social experiments including polarization dynamics and universal basic income effects. GenSim \cite{tang2025gensim} provides a general-purpose platform supporting 100K+ agents with error-correction mechanisms.

Alongside simulation environments, role-playing language agents have been extensively studied. RoleLLM \cite{wang2024rolellm} benchmarks persona consistency across 100 roles, SimsChat \cite{yang2024simschat} generates multi-turn dialogues for 68 characters defined by traits and aspirations, and SocialBench \cite{chen2024socialbench} evaluates agents at both individual and group levels, finding that individual-level proficiency does not imply group-level competence. These systems demonstrate that LLM agents can participate in coherent social interactions, but most do not systematically ground scenarios in validated psychological taxonomies or examine how personality configurations between interacting agents shape joint outcomes.

\subsection{Personality Expression and Operationalization in LLMs}
\label{sec:rw-personality}

Research on personality in LLMs has progressed along three methodological lines: prompting, training, and activation steering. The prompting approach is most established. PersonaLLM \cite{jiang2024personallm} assigned Big Five configurations to 320 personas and found large effect sizes in self-reported BFI scores, with human evaluators identifying traits at up to 80\% accuracy. Serapio-Garc\'ia et al.~\cite{serapio2025psychometric} tested 18 LLMs with psychometric instruments (IPIP-NEO, BFI), demonstrating that personality can be reliably measured and shaped under specific prompting configurations. Additional evidence shows that LLMs form stable, interpretable Big Five patterns across repeated trials \cite{sorokovikova2024llms}.

More recent work has moved beyond prompting. BIG5-CHAT \cite{li2025big5chat} uses supervised fine-tuning and DPO on a 100K-dialogue dataset grounded in real human personality expressions, outperforming prompt-based methods on psychometric measures. Activation-steering approaches use representation engineering to directly manipulate personality-related internal representations \cite{ong2025cooperative}, finding that higher Agreeableness improves cooperation but increases exploitation vulnerability. However, psychometric measurement remains challenging: PERSIST \cite{tosato2026persist} tests 25 models across 2M+ responses and finds that even 400B+ parameter models show substantial measurement instability under question reordering. Most studies use explicit trait prompting (e.g., \textit{``you are highly agreeable''}) \cite{jiang2024personallm, serapio2025psychometric, sorokovikova2024llms}, which frames personality as an instruction rather than a naturally occurring behavioral tendency. Character-based approaches such as InCharacter \cite{wang2024incharacter} uses psychological interviews of 32 fictional characters and achieves 80.7\% personality alignment with human-perceived types from the Personality Database. Moon~\cite{moon2025binding} develops narrative backstory conditioning that reproduces population-level cooperative behaviors in social dilemmas without explicit trait labels. Our work follows this character-based line, leveraging LLMs' embedded knowledge of well-known movie characters and public figures' behavioral tendencies rather than explicit trait descriptors.

When personality-conditioned agents interact in social tasks, studies show trait effects on cooperative behavior. Huang and Hadfi~\cite{huang2024personality} find that Big Five profiles influence negotiation outcomes and strategy use. NetworkGames \cite{qiu2025networkgames} assigns MBTI types to agents in iterated Prisoner's Dilemma on network topologies, showing that macro-level cooperation depends on both dyadic personality pairings and network structure. Zeng et al.~\cite{zeng2025dynamic} model dynamic personality evolution across evolutionary generations. However, most of these studies examine personality within narrow task domains (e.g., cooperative games) and focus on individual trait expression rather than systematic dyadic composition across diverse social contexts, which is the central focus of our work.

\subsection{Agreeableness and Interpersonal Conflict}
\label{sec:rw-agreeableness}

Among the Big Five dimensions, agreeableness has the strongest theoretical and empirical connection to interpersonal conflict and cooperation. Graziano et al.~\cite{graziano1996perceiving} demonstrated through multi-method designs that low-agreeableness individuals rate power assertion significantly more favorably during conflict than their high-agreeableness counterparts. Jensen-Campbell and Graziano~\cite{jensen2001agreeableness} established through diary studies that agreeableness is the Big Five dimension most closely associated with conflict processes and outcomes, with low-agreeableness individuals using more destructive tactics that predict poorer adjustment. The most comprehensive quantitative review to date, by Wilmot and Ones~\cite{wilmot2022agreeableness}, synthesizes 142 meta-analyses across 275 variables, confirming that agreeableness produces desirable effects for 93\% of variables examined. Thielmann et al.~\cite{thielmann2020personality} provide a complementary theoretical framework identifying situational affordances that moderate personality--prosociality links across economic games.

Recent computational studies converge with these psychological findings. Sakai et al.~\cite{sakai2025effects} test personality steering in repeated Prisoner's Dilemma and find agreeableness is the dominant factor promoting cooperation across multiple LLM models. Noh and Chang~\cite{noh2024llms} report across 1,500 multi-issue negotiation simulations that agreeableness is the most important personality trait for negotiation outcomes.

Our work extends this body of research in three ways. First, we examine the agreeableness effects across diverse social goal categories rather than in a single task domain. Second, we operationalize personality through character-based anchoring rather than explicit trait prompting. Third, we analyze the \textit{behavior strategies} through which agreeableness influences outcomes, i.e., the conversational mechanisms that mediate the personality--outcome relationship.

\section{Methodology}
\label{sec:methodology}
\begin{figure*}[t]
    \centering
    \includegraphics[width=\textwidth]{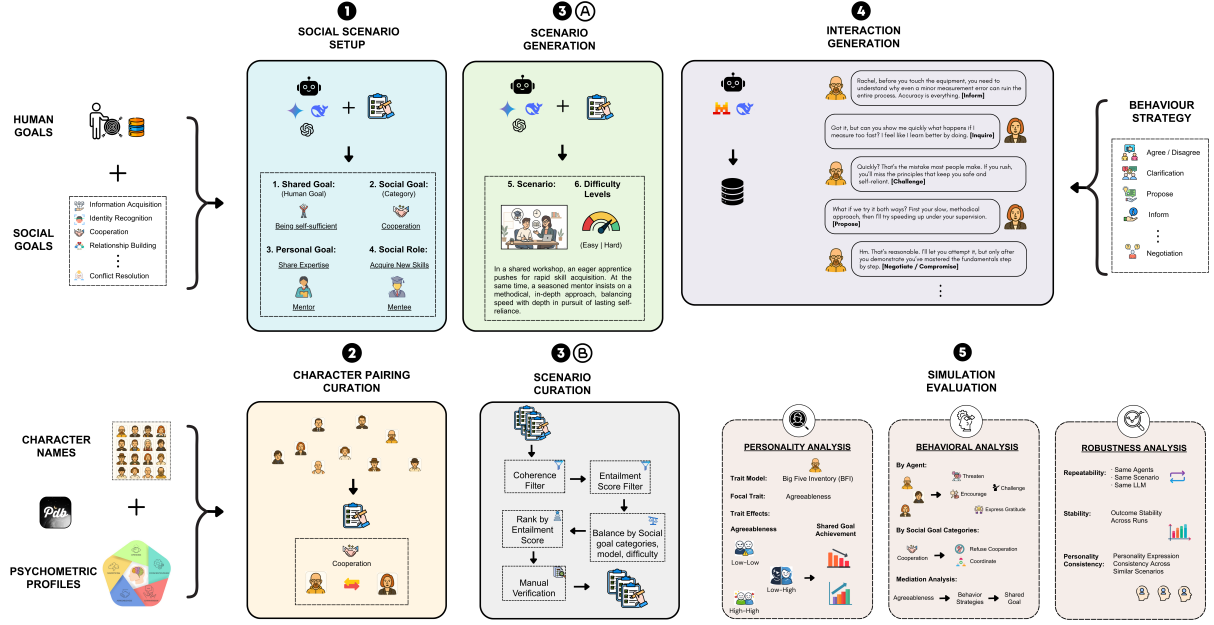}
    \caption{Overview of the simulation pipeline adapted from CHARISMA. \textbf{Stage~1}: Social scenario setup derives shared goals, social goal categories, personal goals, and social roles from a 135-goal taxonomy (See Section~\ref{sec:method-setup}). \textbf{Stage~2}: Characters with crowd-sourced Big Five profiles from the Personality Database are paired into four Agreeableness conditions (See Section~\ref{sec:method-characters}). \textbf{Stage~3}: Scenarios are generated at two difficulty levels and curated through coherence filtering, entailment scoring, balanced selection, and manual verification (See Section~\ref{sec:method-scenarios}). \textbf{Stage~4}: Two LLM agents engage in 20-turn dialogues where each turn involves behavior strategy selection, personality reasoning, and response generation (See Section~\ref{sec:method-interaction}). \textbf{Stage~5}: Evaluation covers personality--outcome analysis, behavioral mediation, and robustness assessment (See Section~\ref{sec:method-evaluation}).}
    \label{fig:framework}
\end{figure*}

We introduce a simulation pipeline adapted from the CHARISMA framework \cite{sadirijavadi2026charisma} for a large-scale empirical study of how dispositional traits and situational factors jointly shape social behavior in social interactions. As shown in Figure~\ref{fig:framework}, the simulation pipeline consists of five stages: \textit{(1)}~social scenario setup, \textit{(2)}~character pairing curation, \textit{(3)}~scenario generation and curation, \textit{(4)}~interaction generation with behavior strategy, and \textit{(5)}~simulation evaluation.

\subsection{Social Scenario Setup}
\label{sec:method-setup}
Social scenarios are grounded in a validated goal-driven structure. Instead of generating scenarios ad hoc, the framework adopts the 135 Human Goals Taxonomy \cite{chulef2001hierarchical}, a systematically organized hierarchy derived from extensive empirical research. Each goal serves as a seed for scenario generation, providing the shared objective around which agent interactions are structured. To bridge abstract human goals and concrete social interaction patterns, each goal is classified into one of eight social goal categories: \textit{Cooperation}, \textit{Conflict Resolution}, \textit{Relationship Building}, \textit{Relationship Maintenance}, \textit{Identity Recognition}, \textit{Information Acquisition}, \textit{Information Provision}, and \textit{Competition}. These categories are informed by classical social interaction typologies \cite{nisbet1970social} and research on interpersonal goals \cite{clark1979interpersonal}. For each scenario, the framework specifies four components: (1)~a \textbf{shared goal} drawn from the taxonomy, (2)~a \textbf{social goal category} classifying the interaction type, (3)~\textbf{personal goals} for each agent that may complement or conflict with the shared goal, and (4)~\textbf{social roles} defining relational positions. This multi-layered goal structure creates conditions for rich interactions by introducing both alignment and potential tension between agents.

\subsection{Character Pairing Curation}
\label{sec:method-characters}
We operationalize personality through \textit{personality anchoring}, rather than assigning abstract trait scores or relying on explicit trait prompting. This approach leverages LLMs’ embedded knowledge of well-known movie characters and public figures to elicit personality-consistent behavior. Characters are sourced from the Personality Database (PDB), a large-scale crowd-sourced platform where users vote on personality traits using multiple frameworks, including the Big Five Inventory (BFI).

Characters undergo a multi-stage filtering process. First, a \textbf{vote-threshold filter} ensures a sufficient number of independent ratings. Second, an \textbf{inter-rater agreement filter} retains only characters with consistent BFI assessments across raters. Third, characters are ranked using an \textbf{Agreeableness score} that quantifies voting support for the assigned Agreeableness level:
\begin{equation}
    \text{Rank} = \frac{c_{\text{main}}}{c_{\text{main}} + c_{\text{other}}}
    \label{eq:rank}
\end{equation}
where $c_{\text{main}}$ is the vote count for the character’s assigned Agreeableness value and $c_{\text{other}}$ is the vote count for the closest alternative. We select five characters at each of four Agreeableness levels $\{0.0, 0.25, 0.75, 1.0\}$, yielding 20 characters in total (See Appendix \ref{app:character_list} for the full list). The neutral level $\{0.5\}$ is excluded because it lacks distinctive behavioral characteristics.

\begin{table}[h]
\centering
\footnotesize 
\setlength{\tabcolsep}{3.5pt} 
\begin{tabular}{lcc}
\hline
\textbf{Condition} & \textbf{Abbrev.} & \textbf{Agent A \&  B} \\
\hline
Homogeneous-Disagreeable & HoD & $\{0.0, 0.25\}$ \\
Heterogeneous-Extreme    & HeE & $\{0.0, 1.0\}$ \\
Heterogeneous-Moderate   & HeM & $\{0.25, 0.75\}$ \\
Homogeneous-Agreeable    & HoA & $\{0.75, 1.0\}$ \\
\hline
\end{tabular}
\caption{\label{tab:dyadic_conditions} Dyadic personality composition conditions based on Agreeableness levels.}
\end{table}

As shown in Table~\ref{tab:dyadic_conditions}, characters are paired into four conditions that vary in \textit{dyadic personality composition}. 
\textbf{Homogeneous-Disagreeable (HoD)} pairs two low-Agreeableness agents, whereas 
\textbf{Homogeneous-Agreeable (HoA)} pairs two high-Agreeableness agents. 
\textbf{Heterogeneous-Extreme (HeE)} represents the maximum contrast between low and high Agreeableness, and 
\textbf{Heterogeneous-Moderate (HeM)} introduces a moderate contrast between the two agents. 
Within each condition, every agent interacts with every agent from the corresponding level ($5 \times 5 = 25$ pairs), yielding 100 unique dyads across all four conditions.

\subsection{Scenario Generation and Curation}
\label{sec:method-scenarios}

Each scenario setup is expanded into a detailed narrative description at two difficulty levels: \textit{Easy} (straightforward dynamics) and \textit{Hard} (high tension), enabling analysis of how personality effects vary under different situational demands. Three LLMs (\textit{DeepSeek-Chat-v3-0324}, \textit{Google Gemini 3 Flash}, and \textit{OpenAI GPT-5.2}) generate scenario setups from the 135 human goals, with a model-consistency approach ensuring that the same model expands its own setups into full scenarios. Generated scenarios undergo a multi-stage curation pipeline. A \textbf{coherence filter} (threshold $\geq 0.8$), based on LLM-as-a-judge evaluation \cite{liu2023geval}, retains only logically consistent scenarios. An \textbf{entailment filter} (threshold $\geq 0.6$) uses a pre-trained NLI model \cite{lewis2020bart} to verify alignment between scenario specifications and generated content. A \textbf{balanced selection} phase caps scenarios per social goal category while ensuring representation across models and difficulty levels. Finally, a \textbf{manual verification} audit on a stratified 12.6\% subset confirms quality along 5 dimensions: goal clarity, role plausibility, social realism, difficulty alignment, and conversational achievability.

The final curated dataset comprises 277 high-quality scenarios. Of the original 8 social goal categories, \textit{Information Provision} was excluded due to insufficient post-filtering representation, leaving 7 categories that are approximately balanced across models and difficulty levels.

\subsection{Interaction Generation}
\label{sec:method-interaction}

Two LLM-based agents interact in maximum 20-turn dialogues. Consistent with the personality anchoring approach, each agent is instantiated with its character identity, relying on the LLM's internal knowledge of the character rather than providing explicit personality information, along with the scenario context, including its assigned role, personal goal, and shared goal.

A central feature of the interaction protocol is the integration of \textbf{behavior strategy} into the generation process. Rather than generating free-form utterances, each agent follows a structured turn-taking sequence:

\begin{enumerate}[leftmargin=*]

    \item \textbf{Behavior strategy selection}: the agent selects a communicative intent label (e.g., \textit{Propose}, \textit{Challenge}, \textit{Encourage}) from a coding scheme organized into category-specific and universal codes (See Appendix \ref{app:behavior_strategies} for the full list). It can also select \textit{None} if no code fits the response.
    \item \textbf{Personality reasoning}: the agent reasons about how its personality traits should influence the response.
    \item \textbf{Response generation}: guided by the selected code and personality reasoning, the agent produces a natural-language utterance. \item \textbf{Trait score reporting}: the agent reports numerical BFI scores reflecting trait levels expressed in the current turn.
    \end{enumerate}

Individual behavior strategies are aggregated into three higher-order \textbf{behavior strategy groups}: \textit{Cooperative} (e.g., Encourage, Express Gratitude, Build Consensus), \textit{Confrontational} (e.g., Challenge, Dismiss, Taunt, Threaten), and \textit{Neutral} (e.g., Inquire, Clarify, Inform). This aggregation enables analysis of how Agreeableness configurations relate to conversational strategy selection and, subsequently, to interaction outcomes. This provides the analytical basis for the behavioral mediation analysis described in Section~\ref{sec:exp2}.

\subsection{Simulation Evaluation}
\label{sec:method-evaluation}

Evaluation covers three complementary dimensions, corresponding to our experimental research questions.

\noindent\textbf{Personality--Goal Achievement analysis} examines how shared goal achievement scores vary across the four Agreeableness pairing conditions. Goal achievement is assessed using an LLM-as-a-judge approach: the evaluator model receives the complete interaction transcript, scenario specification, and scoring rubric, then assigns scores on a 0--10 scale for both shared and personal goal achievement with accompanying reasoning and confidence assessments.

\noindent\textbf{Behavioral mediation analysis} examines how conversational strategy distributions differ across Agreeableness conditions and social goal categories, and whether conversational strategies mediate the relationship between dyadic Agreeableness composition and goal achievement. This analysis operates at multiple levels: pairing-condition aggregates, and mediation pathways linking personality $\rightarrow$ conversational strategies $\rightarrow$ shared goal achievement.

\noindent\textbf{Robustness analysis} assesses two forms of reliability. \textit{Results consistency} is measured by repeating simulations under identical conditions (same agent pair, scenario, and LLM) across multiple runs and computing intraclass correlation coefficients (ICC). \textit{Personality expression consistency} evaluates whether the same character exhibits stable Agreeableness expression across different scenarios within the same social goal category, testing a core assumption of personality anchoring: that LLM agents can embody stable personality profiles through character knowledge alone.
\section{Experiments and Results}
\label{sec:experiments-results}
We conduct four experiments across 1,010 conversations to examine how dyadic Agreeableness composition shapes social interaction outcomes, the behavior strategies underlying this relationship, and the robustness of both results (i.e., shared goal achievement) and personality expression. Table~\ref{tab:experiments} summarizes the experimental design.

\begin{table}[h]
\centering
\small
\setlength{\tabcolsep}{3pt} 
\begin{tabular}{lrc}
\hline
\textbf{Research Question} & \textbf{\# Conv.} & \textbf{Focus} \\
\hline
1. Personality $\rightarrow$ GA & 400 & Direct effects \\
2. Personality $\rightarrow$ BS $\rightarrow$ GA & 400 & Mediation \\
3. Result Consistency & 250 & Robustness \\
4. Personality Expression & 360 & Trait stability \\
\hline
\end{tabular}
\caption{\label{tab:experiments} Overview of the experimental design, including the number of conversations and the analytical focus for each RQ. GA = Goal Achievement; BS = Behavior Strategy. Experiments 1 and 2 are conducted on the same conversation dataset.}
\end{table}

\paragraph{Shared Configuration.} All experiments build on the curated scenario database of 277 scenarios spanning 7 social goal categories. The primary interaction model is \textit{DeepSeek-Chat-v3-0324}, with \textit{Mistral Large} as a cross-model replication. Each conversation comprises 20 turns (10 per agent). Evaluation uses \textit{DeepSeek-Chat-v3-0324} as an LLM-as-a-judge, scoring shared and personal goal achievement on a 0--10 scale with accompanying reasoning and confidence assessments.

\subsection{Experiment 1: Personality and Goal Achievement}
\label{sec:exp1}

\paragraph{Design.} We generate 400 conversations distributed evenly across the four Agreeableness pairing conditions (100 per condition). Each of the 100 unique agent pairs participates in 4 conversations, each assigned to a distinct social goal category via constrained randomization ensuring balance across the 7 categories ($\sim$57 conversations per category).

\paragraph{Results.} Table~\ref{tab:goal_achievement} presents the primary outcome measure: mean shared goal achievement by pairing condition. Shared goal achievement increases monotonically from HoD to HoA, with a 5-point spread on the 10-point scale.

\begin{table}[h]
\centering
\small
\setlength{\tabcolsep}{3pt}
\begin{tabular}{llcc}
\hline
\textbf{Pair Type} & \textbf{Agreeableness} & \textbf{Mean} & \textbf{Success@8} \\ 
\hline
HoD & 0.0 -- 0.25 & \textbf{2.3} & \textbf{6\%} \\
HeE & 0.0 -- 1.0  & 3.7 & 11\% \\
HeM & 0.25 -- 0.75 & 5.6 & 38\% \\
HoA & 0.75 -- 1.0  & \textbf{7.3} & \textbf{62\%} \\
\hline
\end{tabular}
\caption{\label{tab:goal_achievement} Mean shared goal achievement (0--10) and strong success rate (score $\geq 8$) by Agreeableness pair type.}
\end{table}

Applying a threshold of shared goal score $\geq 8$ (``strong success''), HoA pairs achieve strong success in 62\% of interactions compared to only 6\% for HoD pairs, showing a significant difference. Mixed pairs show intermediate success rates but remain closer to HoD than HoA, indicating that consistently high-level success is primarily associated with mutually high Agreeableness rather than the presence of a single agreeable agent.

\begin{figure}[t]
    \centering
    \includegraphics[width=\columnwidth]{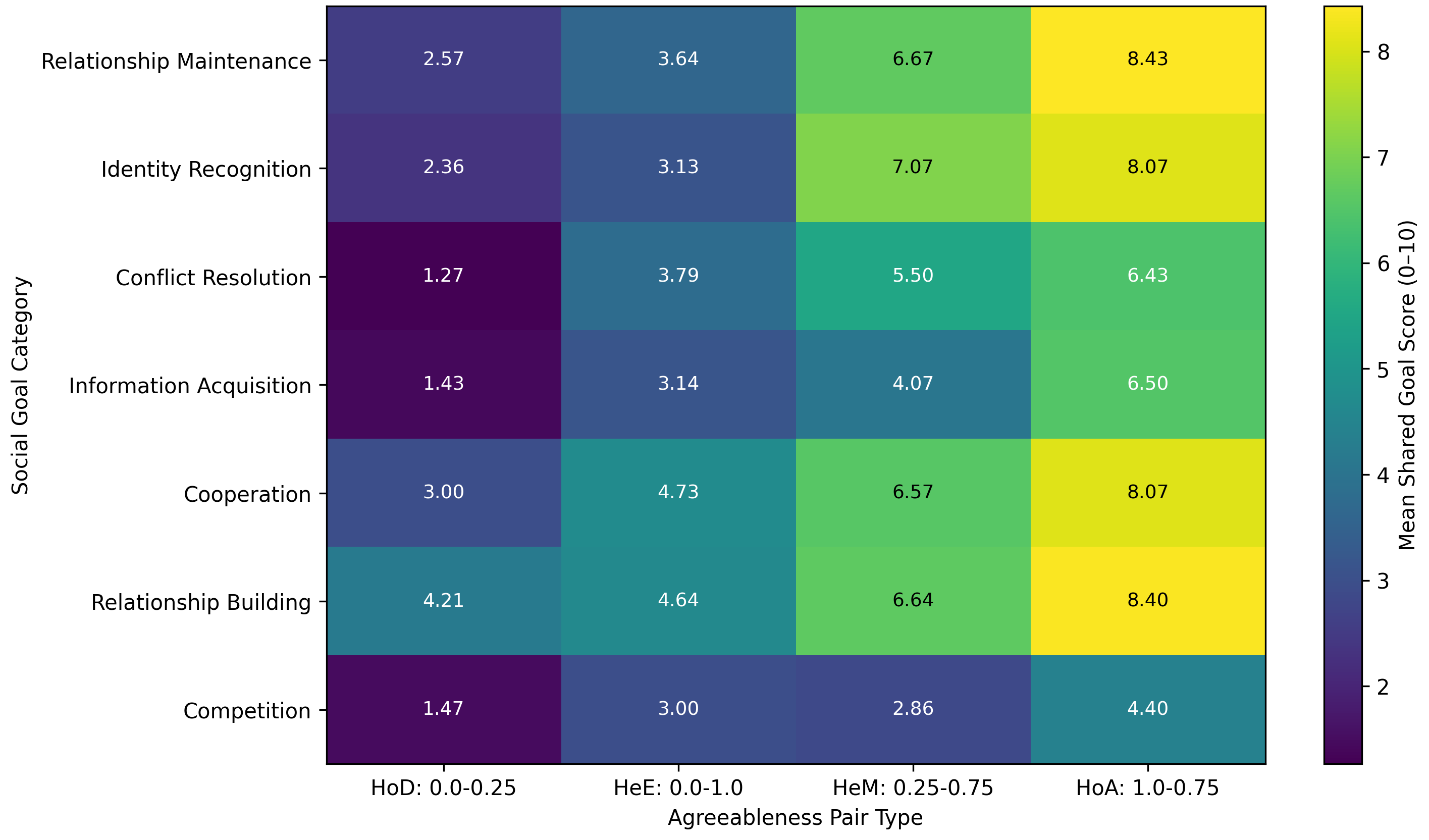}
    \caption{Mean shared goal achievement by social goal category $\times$ Agreeableness pair type. The HoD $<$ HoA contrast is preserved across all categories, with the largest effects in relationally oriented categories and the smallest in Competition.}
    \label{fig:heatmap}
\end{figure}

As shown in Figure~\ref{fig:heatmap}, the Agreeableness effect holds across all seven social goal categories. Categories such as Relationship Maintenance, Identity Recognition, and Cooperation show the strongest HoD--HoA contrast, while Competition shows the smallest effect (HoD: 1.5; HoA: 4.4), suggesting that competitive scenarios pose structural challenges that high Agreeableness alone cannot fully overcome. The effect holds across both difficulty levels (Easy: HoD 2.2, HoA 7.5; Hard: HoD 2.4, HoA 6.9) and all three scenario-generating models (DeepSeek: HoD 2.1, HoA 7.5; Gemini: HoD 2.1, HoA 6.2; OpenAI: HoD 2.8, HoA 8.0), confirming that the observed pattern is not an artifact of specific scenario sources or ceiling effects.

\paragraph{Cross-model replication.} Replicating with Mistral as the interaction backbone preserves the same monotonic ordering and comparable success rates (HoD: 4\%; HoA: 61\%), confirming that the personality--outcome relationship generalizes across interaction models.

\subsection{Experiment 2: Behavioral Analysis}
\label{sec:exp2}

\paragraph{Design.} Using the same 400 conversations from Experiment~1, we analyze behavior code distributions across pairing conditions and examine whether conversational strategies mediate the Agreeableness--outcome relationship.

\begin{figure}[t]
    \centering
    \includegraphics[width=\columnwidth]{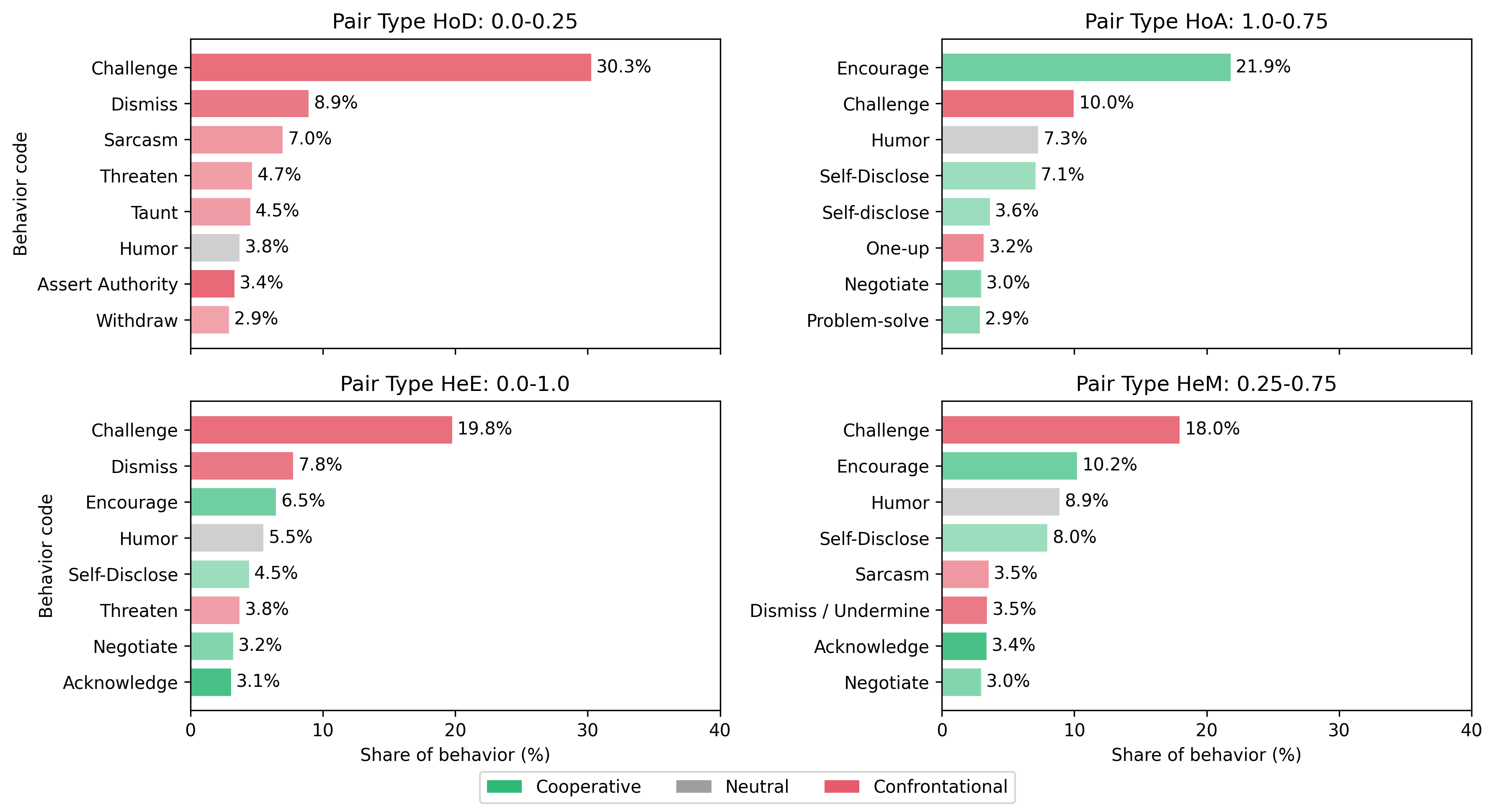}
    \caption{Top behavior strategies by Agreeableness pair type. HoD pairs are dominated by confrontational behaviors (Challenge, Dismiss), while HoA pairs favor cooperative strategies (Encourage, Express Gratitude, Build Consensus).}
    \label{fig:behavior_profiles}
\end{figure}

\paragraph{Behavioral profiles by pair type.} As illustrated in Figure~\ref{fig:behavior_profiles}, Agreeableness configuration produces distinct behavioral signatures. HoD pairs are dominated by confrontational behaviors: \textit{Challenge} (17.3\%), \textit{Dismiss}, \textit{Taunt}, and \textit{Withdraw} characterize the interaction style. HoA pairs exhibit a predominantly prosocial profile: \textit{Humor} (24.2\%) and \textit{Encourage} (17.3\%) together account for over 40\% of behaviors, supplemented by \textit{Express Gratitude}, \textit{Build Consensus}, and \textit{Self-Disclose}. Notably, \textit{Encourage} is virtually absent in HoD pairs, while \textit{Challenge} drops from 17.3\% (HoD) to 4.9\% (HoA). Mixed pairs exhibit intermediate profiles reflecting both orientations.

\begin{figure}[t]
    \centering
    \includegraphics[width=\columnwidth]{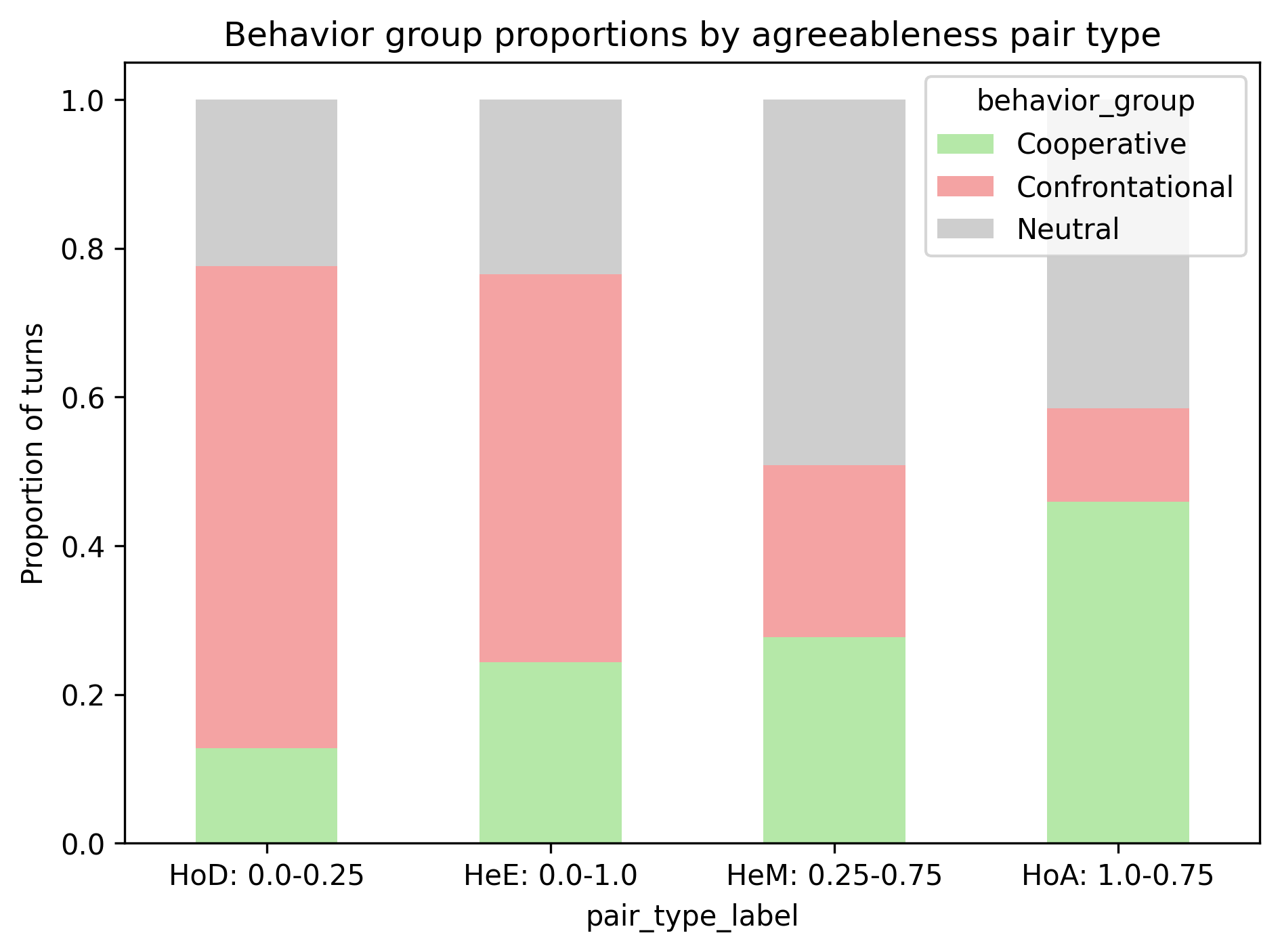}
    \caption{Behavior group proportions by Agreeableness pair type. HoD pairs employ predominantly Confrontational strategies, while HoA pairs favor Cooperative and Neutral behaviors.}
    \label{fig:behavior_proportions}
\end{figure}

\paragraph{Behavior strategy groups and goal achievement.} As shown in Figure~\ref{fig:behavior_proportions}, when individual behavior strategies are aggregated into Cooperative, Confrontational, and Neutral strategy groups, the mediation pathway becomes clear. HoD pairs employ predominantly Confrontational strategies ($\sim$55\% of turns) with minimal Cooperative behavior ($\sim$20\%), while HoA pairs show the reverse pattern. Conversations dominated by Cooperative strategies achieve substantially higher shared goal scores (mean = 7.1) compared to Neutral (mean = 5.3) or Confrontational-dominant conversations (mean = 2.5).

\paragraph{Partial mediation.} As shown in Table~\ref{tab:mediation}, a joint analysis crossing dominant behavior strategy with Agreeableness pair type reveals that Agreeableness continues to predict outcomes \textit{within} the same dominant strategy. When both HoD and HoA pairs employ Cooperative strategies, HoA pairs still outperform HoD (8.1 vs.\ 5.7). Similarly, when both employ Confrontational strategies, HoA pairs achieve higher scores than HoD (3.8 vs.\ 1.6). This pattern indicates \textit{partial} rather than full mediation: Agreeableness shapes outcomes both by increasing the likelihood of cooperative behavior \textit{and} through additional pathways beyond strategy selection.

\begin{table}[h]
\centering
\footnotesize 
\setlength{\tabcolsep}{3pt} 
\begin{tabular}{lccc}
\hline
 & \multicolumn{3}{c}{\textbf{Dominant Behavior Strategy}} \\
\cline{2-4}
\textbf{Pair Type} & Cooperative & Neutral & Confrontational \\
\hline
HoD & 5.7 & 3.3 & 1.6 \\
HeE & 6.5 & 3.8 & 1.8 \\
HeM & 7.0 & 5.2 & 2.3 \\
HoA & 8.1 & 6.2 & 3.8 \\
\hline
\end{tabular}
\caption{\label{tab:mediation} Mean shared goal achievement by strategy $\times$ pair type. Agreeableness predicts outcomes within strategy groups.}
\end{table}

\subsection{Experiment 3: Result Robustness}
\label{sec:exp3}

\paragraph{Design.} Fifty conversation configurations from Experiment~1 are each repeated 5 times under identical conditions (same agent pair, scenario, and model), yielding 250 conversations. Only the stochastic sampling inherent in LLM generation varies across repetitions.

\paragraph{Results.} The pooled standard deviation across repeated runs is 0.98 for shared goal achievement on the 0–10 scale, which is modest relative to the 5-point spread between HoD and HoA means. Single-run intraclass correlation coefficients (ICC\textsubscript{3,1}) reach 0.89 for shared goal achievement, indicating good-to-excellent reliability \cite{koo2016guideline}. When averaged over 5 runs, reliability increases to ICC3\textsubscript{, k} = 0.97. Approximately 94--96\% of configurations exhibit variance below 3.0, confirming that the consistency of results is broadly uniform rather than driven by a subset of stable cases.

\subsection{Experiment 4: Personality Expression Stability}
\label{sec:exp4}

\paragraph{Design.} Each of the 20 agents interacts with 3 partners across 6 scenarios within each of two social goal categories (Cooperation and Conflict Resolution), yielding 360 conversations. A partner balance rule ensures each agent is exposed to both low- and high-Agreeableness partners.

\begin{figure}[t]
    \centering
    \includegraphics[width=\columnwidth]{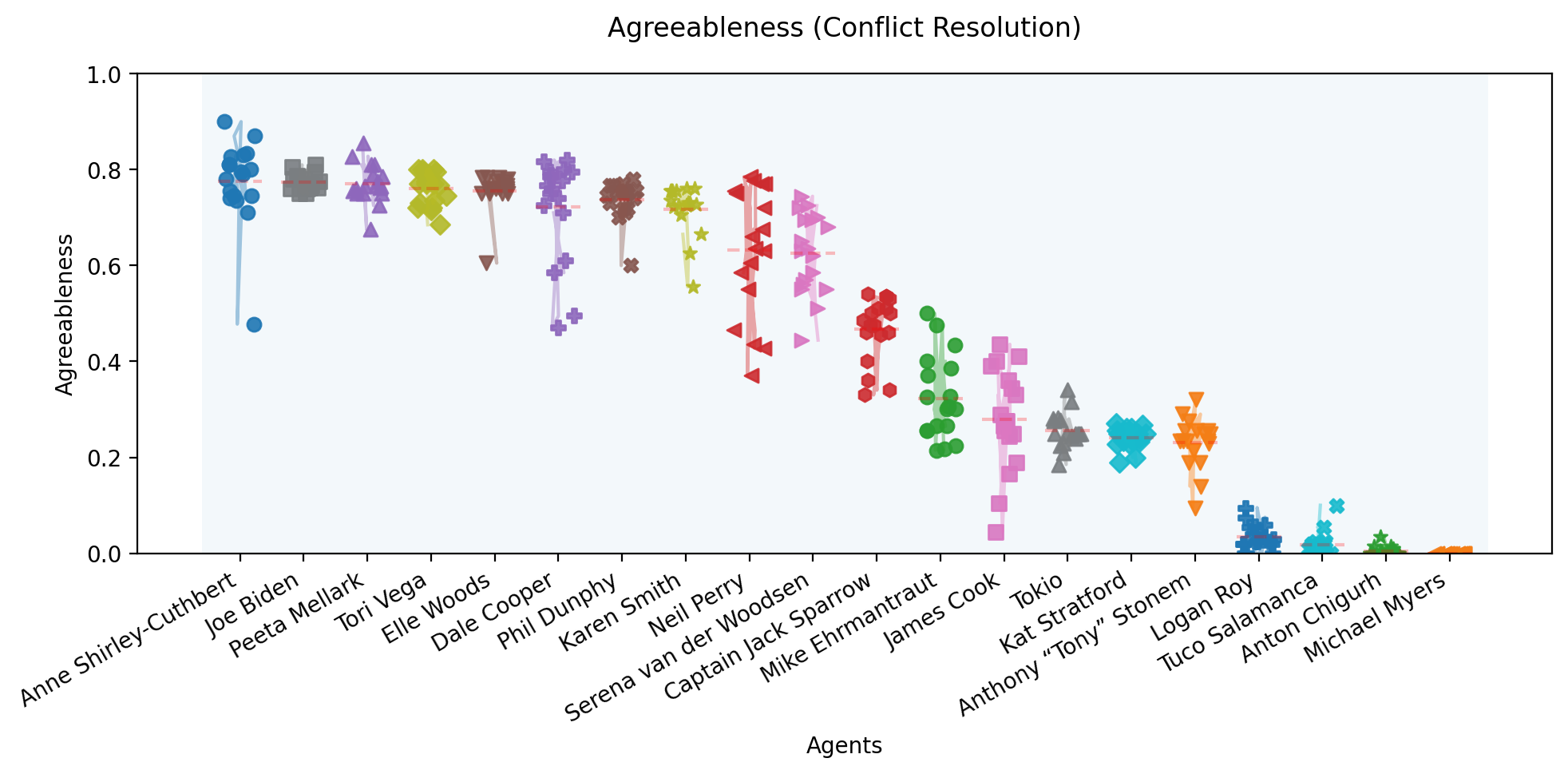}
    \includegraphics[width=\columnwidth]{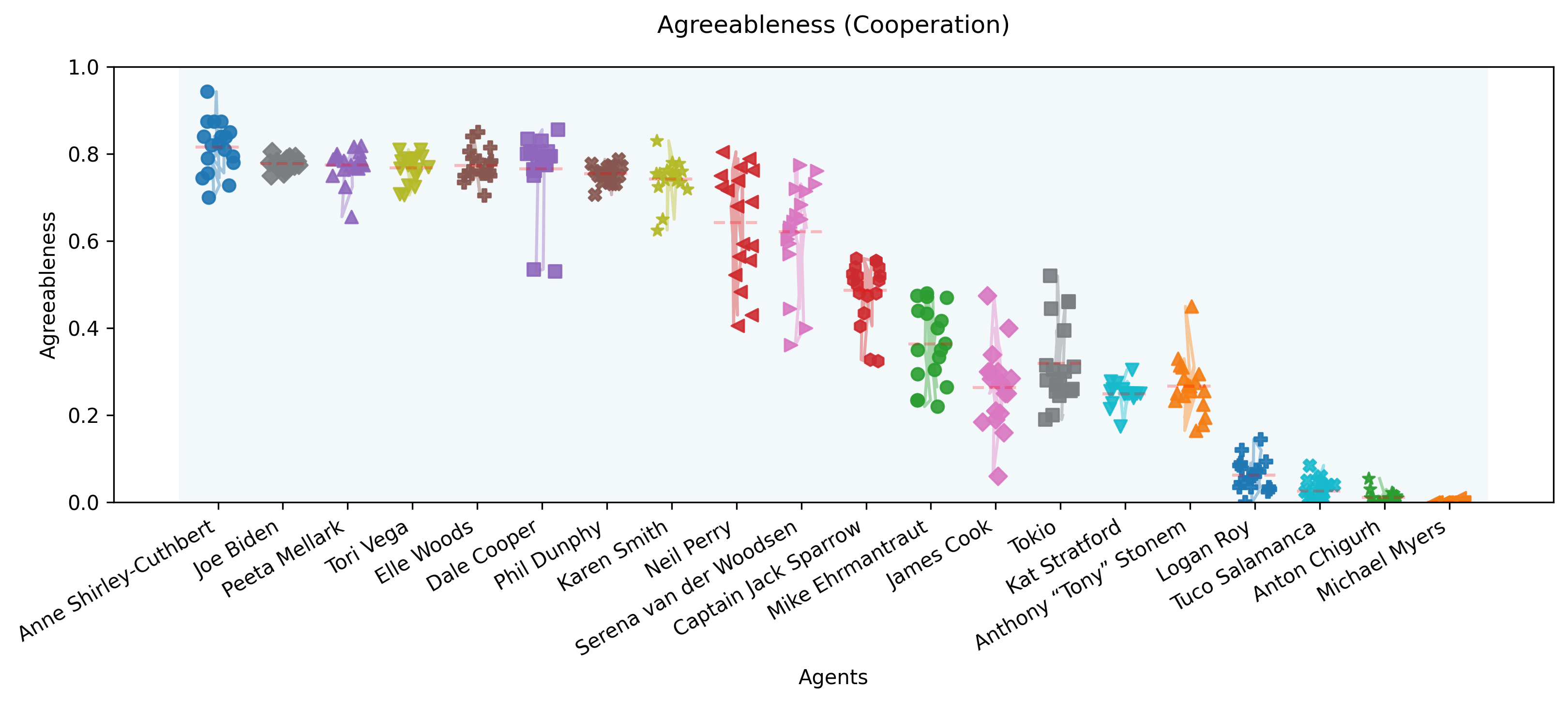}
    \caption{Expressed Agreeableness per agent across scenarios for Conflict Resolution (top) and Cooperation (bottom). Each point represents one conversation. Characters at the extremes show tight clustering; the categorical distinction between low ($<$0.5) and high ($>$0.5) Agreeableness is preserved across all agents.}
    \label{fig:expression}
\end{figure}

\paragraph{Results.} As shown in Figure~\ref{fig:expression}, agents maintain consistent Agreeableness expression across scenarios within each category (Figure~\ref{fig:expression}). Characters at the extremes, such as Logan Roy (expected: 0.0) and Anne Shirley-Cuthbert (expected: 1.0), show tight clustering of expressed values across all scenarios.

More moderate characters show wider but still bounded variability. Critically, the categorical distinction between low and high Agreeableness is preserved across all agents: those expected to be low consistently express values below the 0.5 midpoint, while those expected to be high consistently express values above it.

Expression patterns are similar across the Cooperation and Conflict Resolution categories, suggesting that consistency is a property of the character rather than the situational context. These findings validate personality anchoring as a viable operationalization strategy: without receiving explicit personality scores, LLM agents embody characters in ways that reflect expected trait levels based on the model's internal knowledge.

\section{Conclusion} 
\label{sec:conclusion}

In this paper, we present a large-scale empirical study of how dyadic personality composition shapes social interaction outcomes in LLM-based simulations, using a simulation pipeline adapted from the CHARISMA framework. By leveraging LLMs' embedded knowledge of well-known movie characters and public figures, we operationalize personality as a naturally occurring behavioral tendency grounded in character identity. Across 1,010 simulated conversations spanning seven social goal categories, our findings reveal that first, dyadic Agreeableness composition exhibits a strong monotonic relationship with shared goal achievement. Second, behavioral mediation analysis demonstrates that Agreeableness influences outcomes partially through the selection of cooperative versus confrontational conversational strategies. Third, robustness analyses confirm both high outcome consistency across repeated simulations and stable personality expression across diverse scenarios. By connecting established psychological constructs with LLM-based agent interactions, this work contributes toward a methodological bridge between social psychology and NLP, enabling the systematic examination of how individual differences shape social behavior at a scale.

\section{Limitations} 
\label{sec:limitations}
Several limitations should be acknowledged when interpreting our findings.

\noindent\textbf{1.} Our study focuses exclusively on Agreeableness as the focal personality dimension. While Agreeableness has the strongest theoretical connection to interpersonal conflict and cooperation, social interaction outcomes are likely shaped by the interplay of multiple Big Five traits. Future work should examine how other dimensions, such as Extraversion or Neuroticism, interact with Agreeableness in dyadic settings.

\noindent\textbf{2.} Personality anchoring relies on LLMs' pre-existing knowledge of well-known characters, which introduces potential biases. Characters from Western media dominate the Personality Database, limiting cultural diversity in the agent pool. Moreover, the behavioral tendencies that LLMs associate with specific characters may reflect stereotypical portrayals rather than psychologically nuanced profiles, and these associations may vary across different LLMs depending on their training data.

\noindent\textbf{3.} Our evaluation relies on LLM-as-a-judge scoring for goal achievement assessment. While this approach enables scalable evaluation and has shown alignment with human judgments in prior work, it may introduce systematic biases, for instance, favoring linguistically fluent or explicitly cooperative interactions regardless of actual goal progress. Human evaluation on a larger subset would strengthen the validity of our findings.

\noindent\textbf{4.} Our experiments use a limited set of LLMs for both interaction generation and evaluation. While cross-model replication with Mistral Large provides some evidence of generalizability, the extent to which our findings transfer to other model families remains an open question.
\section{Ethical Consideration} 
\label{sec:ethical-considerations}
Our work raises several ethical considerations that warrant discussion.

\noindent\textbf{1.} Simulating personality-driven social interactions using LLM agents carries the risk of reinforcing stereotypical associations between personality traits and behavioral outcomes. Our finding that low-Agreeableness agents consistently underperform in shared goal achievement should not be interpreted as a deterministic claim about individuals with low Agreeableness in real life, where contextual factors, personal growth, and the multidimensionality of personality play crucial mediating roles.

\noindent\textbf{2.} The use of well-known movie characters and public figures as personality anchors raises questions about representational fairness. Characters are drawn primarily from Western media, which limits the cultural and demographic diversity of the simulated agents. The inclusion of a real political figure (Joe Biden) among the character set further requires caution, as simulated behaviors attributed to real individuals may be misinterpreted as reflecting their actual dispositions or actions.

\noindent\textbf{3.}, while our framework is designed for research purposes in computational social psychology, the methodology could potentially be repurposed to simulate or predict individuals' social behavior based on personality profiles, raising privacy and consent concerns. We emphasize that our work studies aggregate patterns across fictional characters and should not be applied to profile or make judgments about real individuals.

\noindent\textbf{4.} LLM-based social simulation, while offering scalability advantages over traditional experiments, should be understood as a complementary tool rather than a replacement for studies involving human participants. Simulated interactions do not capture the full richness of human social cognition, emotional experience, or moral reasoning, and findings from such simulations should be validated against human behavioral data before informing real-world applications or policy decisions.

\section{Bibliographical References}\label{sec:reference}

\bibliographystyle{lrec2026-natbib}
\bibliography{lrec2026-example}


\clearpage
\onecolumn
\appendix
\section{Appendices}
\label{app:appendices}

\subsection{Character List}
\label{app:character_list}

\begin{table}[h!]
\centering
\scriptsize
\renewcommand{\arraystretch}{1.4}
\begin{tabularx}{\textwidth}{l l X l l c}
\toprule
Character & Category & Subcategory & Genre & Gender & Agreeableness \\
\midrule
Anton Chigurh & Movies & No Country for Old Men (2007) & Crime/Thriller & Male & 0.00 \\
Kat Stratford & Movies & 10 Things I Hate About You (1999) & Romance/Comedy & Female & 0.00 \\
Logan Roy & Television & Succession (2018) & Drama & Male & 0.00 \\
Michael Myers & Movies & Halloween (1978) & Horror & Male & 0.00 \\
Tuco Salamanca & Television & Breaking Bad (2008) & Crime/Drama & Male & 0.00 \\
Anthony “Tony” Stonem & Television & Skins UK (2007) & Drama & Male & 0.25 \\
Captain Jack Sparrow & Movies & Pirates of the Caribbean & Adventure/Fantasy & Male & 0.25 \\
James Cook & Television & Skins UK (2007) & Drama & Male & 0.25 \\
Mike Ehrmantraut & Television & Breaking Bad (2008) & Crime/Drama & Male & 0.25 \\
Tokio & Television & Money Heist (La Casa de Papel) (2017) & Crime/Thriller & Female & 0.25 \\
Anne Shirley-Cuthbert & Television & Anne with an E (2017) & Drama & Female & 0.75 \\
Joe Biden & Political & Presidents of the USA & Political & Male & 0.75 \\
Karen Smith & Movies & Mean Girls (2004) & Comedy & Female & 0.75 \\
Serena van der Woodsen & Television & Gossip Girl (2007) & Drama & Female & 0.75 \\
Tori Vega & Television & Victorious (2010) & Comedy & Female & 0.75 \\
Dale Cooper & Television & Twin Peaks (1990) & Mystery/Drama & Male & 1.00 \\
Elle Woods & Movies & Legally Blonde (2001) & Comedy & Female & 1.00 \\
Neil Perry & Movies & Dead Poets Society (1989) & Drama & Male & 1.00 \\
Peeta Mellark & Movies & The Hunger Games (Franchise) & Science Fiction & Male & 1.00 \\
Phil Dunphy & Television & Modern Family (2009) & Comedy & Male & 1.00 \\
\bottomrule
\end{tabularx}
\caption{List of characters selected for the simulation experiments, including their media category, subcategory, genre, gender, and Agreeableness scores.}
\label{tab:character_pool}
\end{table}

\subsection{Behavioral Coding Scheme}
\label{app:behavior_strategies}
{\scriptsize
\setlength{\tabcolsep}{4pt}
\renewcommand{\arraystretch}{1.3}
\begin{longtable}{l l l p{5.0cm} l}
\caption{Behavioral codebook used for interaction annotation. Each code is associated with a social goal category (Type of Act) and classified into a behavioral strategy group.}
\label{tab:behavior_strategies}\\
\toprule
\textbf{Behaviour Strategy} & \textbf{Definition} & \textbf{Type of Act} & \textbf{Example} & \textbf{Behavioral Group} \\
\midrule
\endfirsthead
\toprule
\textbf{Behaviour Strategy} & \textbf{Definition} & \textbf{Type of Act} & \textbf{Example} & \textbf{Behavioral Group} \\
\midrule
\endhead
\bottomrule
\endfoot
Inquire & Ask direct question & Information Acquisition & "Can you explain why this formula works in practice?" & Neutral \\
Clarify & Seek explanation & Information Acquisition & "Do you mean I shouldn't try the experiment independently yet?" & Neutral \\
Probe & Ask deeper detail & Information Acquisition & "What do you mean by calling the reaction unstable?" & Neutral \\
Challenge & Test claim & Information Acquisition & "But how do you know this method is always reliable?" & Confrontational \\
Request Example & Ask for illustration & Information Acquisition & "Can you show me a time this technique failed?" & Cooperative \\
Check Understanding & Verify comprehension & Information Acquisition & "So you're saying small errors can ruin the whole batch, right?" & Neutral \\
Interrupt & Cut in & Information Acquisition & "Just stop and tell me the answer directly!" & Confrontational \\
Badger & Press repeatedly & Information Acquisition & "Why? Why? Why can't it work differently?" & Confrontational \\
Twist Question & Trap question & Information Acquisition & "So you admit your first explanation was wrong?" & Confrontational \\
Inform & Share fact & Information Provision & "You must heat it to 200°C for stability." & Neutral \\
Elaborate & Add detail & Information Provision & "The temperature matters because molecular bonds are more fragile at lower heat." & Neutral \\
Correct & Rectify & Information Provision & "Actually, it's not sodium chloride, it's sodium carbonate." & Neutral \\
Advise & Suggest practice & Information Provision & "I recommend measuring twice before mixing." & Cooperative \\
Warn & Issue caution & Information Provision & "If you rush this step, the mixture could explode." & Neutral \\
Give Example & Illustrate & Information Provision & "It's like baking—too much flour ruins the cake." & Cooperative \\
Dismiss & Reject & Information Provision & "That's not important right now." & Confrontational \\
Over-explain & Patronize & Information Provision & "Clearly you don't get it, so let me dumb it down." & Confrontational \\
Withhold & Omit & Information Provision & "I'll keep the final step to myself for now." & Confrontational \\
Encourage & Motivate & Relationship Building & "You're improving faster than most beginners." & Cooperative \\
Self-disclose & Share vulnerability & Relationship Building & "I used to panic during my first experiments too." & Cooperative \\
Compliment & Affirm ability & Relationship Building & "You're very precise with your measurements." & Cooperative \\
Humor / Banter & Lighten mood & Relationship Building & "If this blows up, at least we'll have fireworks!" & Cooperative \\
Express Gratitude & Appreciate & Relationship Building & "Thanks for double-checking my notes." & Cooperative \\
Show Interest & Attend & Relationship Building & "How did you come up with that idea?" & Cooperative \\
Exclude & Shut out & Relationship Building & "This discussion isn't for you to join." & Confrontational \\
Mock & Tease hostilely & Relationship Building & "Wow, you're a regular Einstein." & Confrontational \\
Ridicule & Humiliate & Relationship Building & "You'll never get this right, you're too slow." & Confrontational \\
Empathize & Validate & Relationship Maintenance & "I know it's stressful, but you're doing fine." & Cooperative \\
Politeness & Respectful phrasing & Relationship Maintenance & "Could you please explain that again?" & Cooperative \\
Encourage & Sustain motivation & Relationship Maintenance & "We're almost there, keep pushing." & Cooperative \\
Check-in & Reassure & Relationship Maintenance & "Are we still on the same page here?" & Cooperative \\
De-escalate & Calm conflict & Relationship Maintenance & "Let's pause before we argue further." & Cooperative \\
Repair Attempt & Restore harmony & Relationship Maintenance & "Sorry if I came across too harsh earlier." & Cooperative \\
Sarcasm & Dismissive humor & Relationship Maintenance & "Oh sure, you're the master chemist now." & Confrontational \\
Stonewall & Withdraw & Relationship Maintenance & ". . . (silence, no response)" & Confrontational \\
Passive-aggressive & Indirect resistance & Relationship Maintenance & "Fine, I'll do it. . . someday." & Confrontational \\
Withdraw & Detach & Relationship Maintenance & "Whatever, do it yourself." & Confrontational \\
Assert Authority & Establish role & Identity Recognition & "I've taught this for 20 years—you need to follow my lead." & Confrontational \\
Defer / Yield & Accept other's role & Identity Recognition & "You're more experienced, so I'll follow you." & Cooperative \\
Acknowledge Expertise & Recognize status & Identity Recognition & "You're clearly skilled at precision." & Cooperative \\
Attribute / Label & Highlight quality & Identity Recognition & "You're a natural problem-solver." & Neutral \\
Defend Identity & Protect image & Identity Recognition & "I might be new, but I'm capable of learning." & Neutral \\
Challenge Status & Question role & Identity Recognition & "Why should you always be in charge?" & Confrontational \\
Dismiss Identity & Undermine & Identity Recognition & "You're not really qualified to lead." & Confrontational \\
Boast Identity & Overclaim & Identity Recognition & "I'm the smartest one here, no doubt." & Neutral \\
Identity Attack & Insult & Identity Recognition & "You're useless as a mentor." & Confrontational \\
Propose & Suggest plan & Cooperation & "I'll measure, you handle mixing." & Cooperative \\
Negotiate & Balance needs & Cooperation & "We can try your method first, then mine." & Cooperative \\
Coordinate & Organize & Cooperation & "You start the timer while I weigh the sample." & Cooperative \\
Assist & Help & Cooperation & "I'll grab the glassware for you." & Cooperative \\
Build Consensus & Align group & Cooperation & "Do we all agree on this approach?" & Cooperative \\
Share Resources & Provide tools & Cooperation & "Here's my notebook—you can use the data." & Cooperative \\
Reluctant Cooperation & Half-hearted & Cooperation & "Fine, I'll do it, but only this once." & Confrontational \\
Conditional Help & Attach strings & Cooperation & "I'll help if you do my task later." & Confrontational \\
Undermine Cooperation & Fake help & Cooperation & "I'll mix this—oops, spilled it." & Confrontational \\
Refuse Cooperation & Deny & Cooperation & "No, I won't work with you on this." & Confrontational \\
Criticize & Express dissatisfaction & Competition & "This is way too slow." & Confrontational \\
Defend & Hold position & Competition & "No, my method is better than yours." & Neutral \\
One-up & Compare & Competition & "I got better results than you did." & Confrontational \\
Claim Credit & Ownership & Competition & "That was my idea, not yours." & Neutral \\
Boast & Self-promotion & Competition & "I'm the fastest in this class." & Neutral \\
Dismiss / Undermine & Belittle & Competition & "Your approach is useless." & Confrontational \\
Sabotage & Obstruct & Competition & "I didn't give you the full instructions." & Confrontational \\
Refuse to Share & Withhold & Competition & "No, I won't tell you my method." & Confrontational \\
Taunt & Intimidate & Competition & "You'll never keep up with me." & Confrontational \\
Exploit Weakness & Attack vulnerability & Competition & "You always panic—this will break you." & Confrontational \\
Persuade & Shift perspective & Conflict Resolution & "Try it my way—it's safer and faster." & Cooperative \\
Mediate & Reframe & Conflict Resolution & "Let's focus on our shared goal instead." & Cooperative \\
Problem-solve & Suggest fix & Conflict Resolution & "What if we combine both approaches?" & Cooperative \\
Concede & Back down & Conflict Resolution & "Alright, we'll do it your way." & Cooperative \\
Acknowledge Fault & Admit & Conflict Resolution & "I was too impatient earlier." & Cooperative \\
Express Regret & Apologize & Conflict Resolution & "I shouldn't have snapped at you." & Cooperative \\
Disagree & Reject purposal & Conflict Resolution & "I can't support that plan." & Neutral \\
Blame & Accuse & Conflict Resolution & "This mistake was your fault." & Confrontational \\
Threaten & Intimidate & Conflict Resolution & "If you ignore me, I'll quit." & Confrontational \\
Escalate & Intensify & Conflict Resolution & "This is ridiculous—I'm done with this team!" & Confrontational \\
Counter-accuse & Deflect blame & Conflict Resolution & "Don't blame me—it was your error." & Confrontational \\
Acknowledge & Recognize statement & Universal & Right, I follow you there. & Cooperative \\
Express Emotion & Show feeling & Universal & That actually frustrates me a bit. & Neutral \\
Humor & Use humor or irony & Universal & Well, that went up in smoke faster than my last plan! & Neutral \\
Self-Disclose & Share experience & Universal & Back when I started, I made the same mistake. & Cooperative \\
Encourage & Sustain motivation & Universal & That’s worth exploring further. & Cooperative \\
Reflect & Restate point & Universal & So you’re saying the deadline’s the real issue. & Neutral \\
Meta-Comment & Note conversation flow & Universal & We seem to be talking past each other right now. & Neutral \\
Challenge & Question idea & Universal & Maybe, but have you considered the downside? & Confrontational \\
Interrupt & Cut in to speak & Universal & Hold on—let me finish that point. & Confrontational \\
Dismiss & Reject input & Universal & That’s not really relevant. & Confrontational \\
Sarcasm & Mock indirectly & Universal & Oh sure, because that worked so well last time. & Confrontational \\
Deflect & Shift topic & Universal & Let’s not get into that right now. & Neutral \\
Shift Topic & Move discussion & Universal & Anyway, about tomorrow’s plan… & Neutral \\
Withdraw & Pull back participation & Universal & I think I’ll stay out of this one. & Confrontational \\
Express Gratitude & Thank contribution & Universal & Thanks, that helps clarify things. & Cooperative \\
\end{longtable}
}

\end{document}